RESEARCH ARTICLE                                                                                                          OPEN ACCESS

# A Survey on Deadline Constrained Workflow Scheduling Algorithms in Cloud Environment

Nallakumar. R[1], Sruthi Priya. K. S[2]
Assistant Professor[1], PG Scholar[2]
Department of Computer Science and Engineering
Anna University Regional Centre, Coimbatore
TamilNadu - India

**ABSTRACT**
Cloud Computing is the latest blooming technology in the era of Computer Science and Information Technology domain. There is an enormous pool of data centres, which are termed as Clouds where the services and associated data are being deployed and users need a constant Internet connection to access them. One of the highlights in Cloud is the delivering of applications or services in an on-demand environment. One of the most promising areas in Cloud scheduling is Scheduling of workflows which is intended to match the request of the user to the appropriate resources. There are several algorithms to automate the workflows in a way to satisfy the Quality of service (QoS) of the user among which deadline is considered as a major criterion, i.e. Satisfying the needs of the user with minimized cost and within the minimum stipulated time. This paper surveys various workflow scheduling algorithms that have a deadline as its criterion.

## I. INTRODUCTION

Cloud Computing is one of the budding domains that has gained popularity in the recent years among all those users who make use of the Internet and also it is an amalgamation of Grid and Utility computing. High reliability, extendibility, versatility and Virtualization mark the characteristics of this upcoming technology. It makes use of a pay-per-use-model to share and deliver huge no of computational resources to the users with the procurement of virtualization technologies. Cloud Computing serves the needs of the users in various business and scientific applications. Use of these also reduces the cost of hardware and software as the resources are being leased [1] only for a particular period of time as demanded by the end user.

A diverse range of research issues is addressed in Cloud namely security, fault tolerance, workflow scheduling. Among these deadlines constrained workflow scheduling is the utmost issue in the era of computing. Various new features rendered by Cloud Computing enables the efficient automation and execution of workflows. Workflow Scheduling, a common NP-complete problem is the process of mapping the workflow tasks to the appropriate resources with the help of efficient scheduling algorithms in a way to satisfy the Quality of Service (QoS) of the user along with cost optimization and deadline constraint.

Workflows structures the applications using a directed acyclic graph (DAGs) where tasks are denoted by the nodes and dependencies between the tasks are represented by edges.

Workflows are widely used in many scientific applications such as Bioinformatics, Astronomy and physics. Scientific Workflows comprise of many multi-stage scientific computations that are represented in a highly detailed way using scripting languages, data flow, Petri nets and directed acyclic graphs (DAGs). The main aim is to focus on the workflows, its need, applications and to emphasize the comparison of various workflow scheduling algorithms that has a deadline as its criterion.

The remainder of the paper is structured as follows: The workflow scheduling, is described in section II followed by the survey on existing workflow scheduling algorithm with deadlines as a criterion in section III. Section IV concludes the paper.

## II. WORKFLOW SCHEDULING

Workflow Scheduling deals with the identification of appropriate resources and allocation of the tasks on those resources. Workflows mainly concentrate on the resource automation of procedures to pass the files and data among the participants based on a set of rules [2]. A workflow represents a set of tasks and the dependencies among the tasks with the help of a directed acyclic graph (DAG) G= ($V_i$, $E_i$) where $V_i$ represents the no of tasks in the workflow and $E_i$ denotes the data dependencies between those tasks. A sequence of logically connected, coherent steps which exhibits parent child relationships [3] had to be followed in a workflow. A set of rules defines the linkage between parent and child tasks.In such a relationship parent task had to be executed before its child task gets executed. Tasks that don't have a parent are termed as Entry tasks and those without a child are called as Exit tasks. Figure1 shows the different tasks present in a workflow. The entry task is task 1 and the exit task is task 6. The child tasks 2 and 3 are executed only after the execution of their parent task namely 1.

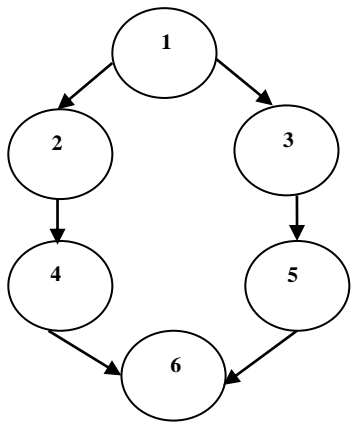

Fig 1 Workflow representation in the form of Directed Acyclic Graphs (DAG)

Workflow Management deals with the scheduling of workflows in a fruitful way. As the performance of the system is greatly influenced by the accurate and correct execution of workflows, we need to choose the right scheduling algorithm that finds the optimal schedule to satisfy the Quality of Service (QoS) constraints demanded by the user such as cost minimization along with deadline and budget constrained, reliability, etc. More concrete examples of workflows include E-business scenarios, online banking, ATM (Automated Teller Machine) accounts card verification in which the user will enter the password and the administrator or the system will verify the password and check the password with the original password. If both the original password, and the one given by the user matches, system will allow the user to access the account, otherwise user access will be denied and that a warning message will be displayed. Figure2 shows the ATM (Automated Teller Machine) account verification process that is represented as a workflow with a series of steps.

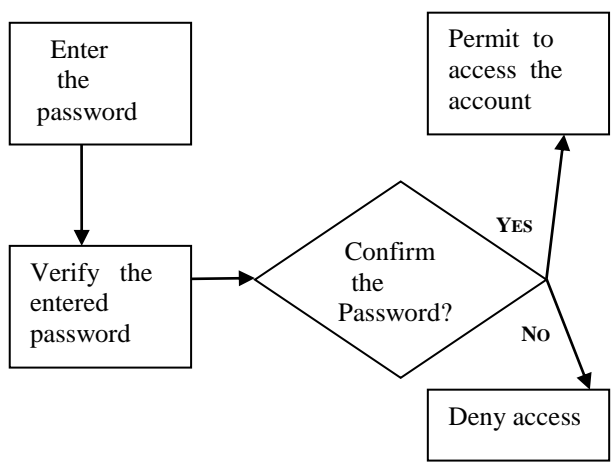

Fig 2 Simple Workflow process for ATM (Automated Teller Machine) account verification.

*A. Necessity for Workflows in Cloud*

Earlier grid computing was used to implement the Workflows. Performance level in grids was very low when compared to that of the Cloud. This was the phenomenon that led to the movement of workflows in the Cloud environment. Cloud computing enabled the provisioning of resources to meet the requirements of the application, which is a guiding factor that enables the Workflow Management System to meet the QoS (quality of the Service) requirements of the application. Grid resources, lack complete environment for execution, which is very vital for the workflow applications. The cloud offers application scalability, high bandwidth, storage and computation are also available at minimal costs and all these together pushes workflow applications to be executed in a cloud environment.

*B. Scientific Workflows*

Workflows that are used to model various aspects of scientific disciplines are named as Scientific Workflows. They have the ability to get parallelized on distributed resources automatically. Cloud computing forecasts a new way of thinking to execute and deploy Scientific Workflows. Using Scientific workflows many complex applications can be broken down into smaller components and can be executed reliably and efficiently. The various benefits that these workflows, offers are [4] (a) enabling the scientists to execute their workflows in easy -to-use environment with the help of interactive tools, the process of sharing the workflows between scientists are simplified, helps the scientists to reproduce previous results, promotes the analyses to be carried out in an infrastructure that is not reliable. Scientific Workflow systems are those that are used to compose and execute a series of workflows in scientific applications.

*C. Examples of Scientific Workflow Systems*

Examples of Scientific Workflow systems [5] include Bioclipse-represent complex actions as workflows in a scripting environment, Orange-Open source data visualizations and analysis, PipeLine Pilot, KNIME, Vis Trails, Traverna workbench - employed in Bioinformatics, ASKALON -a workflow execution for Cloud and grid, Apache Airavata, BioBIKE, Discovery Net: an oldest execution system for workflow, Ergatis, galaxy. The Pegasus Workflow management system is one of the typical examples of scientific workflow application which compiles the complex workflows into executable one. It offers scalability to handle millions of tasks with the huge no of resources. It also provides portability. The other remarkable features of this comprises the following. It enhances performance and scalability by clustering of the tasks, workflow reduction, fault tolerance and workflow reduction.

## III. SURVEY ON CURRENT WORKFLOW SCHEDULING ALGORITHMS WITH DEADLINE AS A CRITERIA

Some of the commonly used algorithms to schedule workflows with deadline as one of its major criterion are Ant Colony Optimization, Genetic algorithms, Partial Critical Path (PCP) algorithm, Particle Swarm Optimization Algorithm (PSO) whose explanation follows below and other algorithms such as Novel DBC (deadline Budget Constrained), Dynamic Provisioning Dynamic Scheduling (DPDS), Fault Tolerant Workflow Scheduling (FTWS) are the ones that are used to schedule workflows that focus on a particular problem domain.

**A.** *Ant Colony optimization Algorithms*

Workflow scheduling using Ant Colony Optimization is explained in by W. N. Chen, J. Zhang and Yang. Y in 2007 [6]. The ant colony optimization algorithm is one of the most vital algorithm to solve computational problems and it uses heuristics. Complex problems are reduced into simpler ones to find out the good optimized path through the graphs. This algorithm is based on the fact of how ants find a path between their colony and the source of food. An ant colony optimization algorithm is a Meta-heuristic one. The implementation of this algorithm involves a no of concurrent steps. The heuristic information has to be initialized first, followed by the generation of ants. Next we need to map the ants with the path and the objective function has to be evaluated.

*B. Genetic Algorithms*

Jia Yu and Rajkumar Buyya explain the use of genetic algorithms to schedule scientific workflows with budget and deadline constraints [7]. Genetic algorithm belongs to the class of evolutionary algorithms. It produces an optimization solution in polynomial time using various evolutionary techniques from a large search space. The fitness function plays a key role in this type of algorithm which determines how much an organism can reproduce before it dies. Genetic algorithms make use of operations like mutation, recombination and selection. Problems which need an optimized solution and that has to be implemented within the specified deadline depends widely on genetic algorithms.

*C. Particle Swarm Optimization (PSO) algorithms*

In the paper [8] the author describes how Particle swarm Optimization is being used efficiently to schedule workflows. This is a population based stochastic optimization algorithm developed by Dr. Eberhart and Dr. Kennedy in 1995 given great ideas from social behavior of bird travelling in (a large group) or fish schooling. PSO shares many homogenous attributes with genetic algorithms, but it doesn't have any evolutionary operators.

The PSO algorithm works by initializing a population (swarm of) candidate solutions (known as particles). Each particle will keep track of its best solution named as personal best(pbest) and also the best value of any particle called as global best(gbest). The particles, fly through the quandary space by following the current optimum particles. Every particle changes its position based on its current position, speed. To add on, the distance between the recent position and pbest, the distance between the current position and gbest are also considered.

*D. Partial Critical Path (PCP) algorithms*

The brief description of the assignment of PCP (partial critical path) is detailed in algorithm one by Rodrigo N. Calheiros and Raj Kumar Buyya in [9]. IC-PCP (IaaS Cloud Partial Critical Path) algorithm schedules the workflow applications in a public cloud environment. It implements the workflows with the criteria of cost minimizations and to meet the deadline specified by the user. This algorithm works by assigning the tasks of the workflow to the Partial Critical Path (PCP). The PCP (Partial Critical Path) will initially hold the parent of one of the exit task which was not already scheduled. If it was already scheduled, the task with the latest finish time will be chosen and added to the Partial Critical Path (PCP).

\

TABLE I
PREVAILING WORKFLOW SCHEDULING ALGORITHMS WITH DEADLINE AS A CRITERION

| S. NO | RELATED WORK | YEAR OF PUBLICATION | SCHEDULING ALGORITHM TYPE | DOMAIN | OVERVIEW |
|---|---|---|---|---|---|
| 1 | Scheduling Workflows using genetic algorithm with a deadline and budget Constraints [7] | 2006 | Genetic algorithm | Grid Computing | (i) A genetic algorithm is used to schedule the workflow applications within the deadline specified by the user.<br><br>(ii) Using Genetic algorithm high-quality solution can be derived from large search space in polynomial time by the use of evolution principle.<br><br>(iii) Target is on heterogeneous and reservation based service-oriented environment to deal with deadline constrained optimization problems. |
| 2 | Workflow scheduling in Grid using Ant colony Optimization [6] | 2007 | Ant Colony Optimization algorithm | Grid Computing | (i) Pheromone and heuristic information is used to guide the search direction of ants to solve bi-criteria problem.<br><br>(ii) One out of each heuristic type and pheromone type are used by each ant in each iteration<br><br>(iii) Workflow scheduling is done efficiently using ACO (Ant Colony Optimization) to complete before the deadline. |
| 3 | Ant Colony Optimization for a Grid scheduling problem [10] | 2009 | Ant Colony Optimization algorithm | Grid computing | (i) Finds the schedule which satisfies all user imposed QoS (Quality of Service) constraints.<br><br>(ii) Pheromone values are calculated based on heuristics and ten workflow applications is used to do experiments |
| 4 | Particle Swarm Optimization for workflow scheduling [8] | 2010 | Particle swarm Optimization algorithm | Cloud Computing | (i) Schedules workflow applications by taking into account the data transmission and computation cost.<br><br>(ii) Workloads are also distributed with |

| | | | | | minimal cost |
|---|---|---|---|---|---|
| | | | | | (iii) PSO (Particle Swarm optimization) gives best results in terms of cost and workload when compared with BRS (Balance Resource selection) algorithm |
| 5 | Fault tolerant workflow scheduling [11] | 2010 | Fault Tolerant Workflow scheduling (FTWS) algorithm | Cloud Computing | (i)Aims to schedule the workflows within the specified deadline, even if failures occur at the time of workflow execution<br><br>(ii)Takes into account the replication and resubmission of tasks to provide good results |
| 6 | Knowledge based Ant Colony Optimization for Grid [12] | 2010 | Ant Colony Optimization algorithm | Grid computing | (i)Knowledge based Ant colony optimization is used.<br><br>(ii)This algorithm finds the workflow schedule with cost minimization as well as deadline as one of the constrain.<br><br>(iii)Pheromone is defined in terms of cost and design and heuristic is defined by the recent starting time of the tasks in workflow applications. |
| 7 | A novel deadline and budget constrained for computational grids [13] | 2011 | Novel DBC (Deadline and Budget Constrained) Algorithm | Grid Computing | (i)Novel DBC (Deadline and Budget Constrained) was used to model sequential workflow application.<br><br>(ii)Tasks are assigned to resources and considers completion time (deadline), budget and relative cost together |
| 8 | Cost and deadline constrained for Scientific Workflows Ensemble in IaaS Clouds [14] | 2012 | Dynamic Provisioning Dynamic Scheduling (DPDS) algorithm | Cloud Computing | (i)Scientific workflows that are grouped into ensembles of workflows are scheduled effectively meeting the deadline constraints.<br><br>(ii)DPDS (Dynamic Provisioning Dynamic Scheduling) is an online algorithm that aims to create ensembles from real time applications<br><br>(iii)CloudSim tool is used to implement this algorithm |

| 9 | Deadline constrained workflows for IaaS [15] | 2013 | PCP (Partial Critical Path) Algorithm | Cloud Computing | (i) PCP is of two types: IC_PCP (IaaS Cloud Partial Critical Path) is executed in one phase and IC-PCPD2 (Infrastructure As A Service cloud partial critical path with deadline distribution) is computed in two phases such as deadline distribution and planning (ii)These two algorithms enable scheduling of larger workflows efficiently as they possess polynomial time complexity. (iii)Executes the workflows in minimum time within the user specified deadline |

## IV. CONCLUSIONS

Workflows have been gaining the main focus in scheduling by moving it from Grid computing to Cloud computing. Related work on the existing workflow scheduling algorithms on deadline constraints have been tabulated based on their domain and type of algorithm which aims to satisfy the Quality of Service (QoS) of the user, such as cost minimization within the particular time limit (deadline) specified by the user. But still there are only limited algorithms that are used generally to automate workflows with deadlines as its criterion such as Ant Colony Optimizations, Genetic algorithms, Particle swarm Optimization (PSO), Partial Critical Path (PCP) algorithms and so there is an urge to explore new algorithms in the near future as a deadline constraint is becoming an important aspect.